\documentclass[twocolumn,showpaces,preprintnumbers,amsmath,amssymb]{revtex4}
\usepackage{amsmath}
\usepackage{graphicx}
\usepackage{color}

\begin{document}

\title{The Gate Voltage Control of Long DNA Coherent
Transport on Insulator Surface}

\author{Zhi-Jie Qin$^1$}\email{qin_zhijie@163.com}
\author{Le Wang$^{1}$}
\author{Gui-Ping Zhang$^2$}

\affiliation{$^1$School of Physics and Engineering, Zhengzhou
University, Zhengzhou, 450001, China}

\affiliation{$^2$Department of Physics, Renmin University of
China, Beijing 100872, China}

\date{\today}

\begin{abstract}

  We investigate the coherent transport properties of a DNA chain on
a substrate which is subjected to a uniform electric field
perpendicular to the surface. On the basis of the effective
tight-binding model which simulates charge transport through DNA,
the transmission coefficient, Lyapunov exponent, and localization
length are numerically calculated by using the transfer-matrix
method. It is found that an isolated extended state may appear at
the Fermi level for a certain gate voltage when the interaction
strength between the chain and the substrate is position dependent
but independent of the base-pair sequence, leading to the gate
voltage induced Metal-insulator transition (MIT). Moreover,
conductance and current-voltage characteristics are also
calculated. The relationship of Lyapunov exponent distribution to
the current-voltage characteristics is discussed. Two different
conduction mechanisms are proposed depending on effectively
delocalized states and isolated extended states, respectively.
These results may provide perspectives for experimental work aimed
at controlling charge transport through DNA-based nanodevices.

\end{abstract}


\maketitle

\section{Introduction}

Since Eley and Spivey \cite{r1} predicted that DNA chain could be
conductive, various ingenious experiments have been designed to
measure the conductance of DNA. Some of them assert that DNA is an
insulator \cite{r2,r3}, but others show the semiconductor or
conductor properties \cite{r4,r5,r6}, and even find the evidence
of its superconductor behavior \cite{r7}. Although controversy
remains for these consequences, it is believed that DNA molecule
has the potential for conducting, whose conductivity not only
depends on its nature, e.g. the sequence, the sugar-phosphate
backbone, and the helix configuration, but also suffers crucial
impacts of the environment, e.g. temperature, humidity, solvent,
ions, substrate and external field\cite{r8,r9}. In 2001, K. H. Yoo
et.al have already found in their experiment that the current
though DNA can be altered by the gate voltage, that is, DNA
molecule is analogous to the field effect transistor(FET)
\cite{r10}. This phenomenon has reappeared several times in the
later experiments \cite{r11,r12,r13,r14}. Theoretically, A.V.
Malyshev attributes this phenomenon to the impact of the external
field on the helix structure \cite{r15}. In his model, the DNA
chain is suspended without any substrate. However, DNA chains are
usually deposited on the surface of some substrate in most of the
experiments. In fact, several groups have observed the strong
interaction between the mica/SiO2 surface and the DNA chain on it,
which does have significant impact on the charge transport through
DNA\cite{r16,r17,r18,r19,r20}. So the study of disorder and
substrate effect in one-dimensional DNA chain is still an engaging
topic. In fact, with the coexistence of the substrate, gate
voltage, and disorder, a complicated situation may appear in DNA
chain.

In the present paper, we consider a one-dimensional(1D)
tight-binding model to investigate the combined effect of
disorder, the interaction between DNA chain and substrate, and the
gate voltage. The disorder is introduced by random distribution of
poly(dA)-poly(dT) and poly(dG)-poly(dC) in DNA chain. An isolated
extended state has been founded for a certain gate voltage when
the interaction strength is dependent on the position of chain but
independent of the base-pair sequence, which leads to the gate
voltage induced Metal-insulator transition (MIT). The dependance
of the transmission spectrum and Lyapunov exponent(LE) on the
energy, the gate voltage and chain length are calculated. The
associated behavior is studied and the relationship of the results
to the distribution of Lyapunov exponent is also discussed.

The paper is organized as follows: In the next section we describe
the basic formalism in our calculations. In section 3 we present
the main results and discuss the related physical implication. The
last section is devoted to a brief summary of conclusions.

\section{THE BASIC FORMALISM}

We consider a 1D DNA chain with interaction between chain and
substrate as illustrated in Fig. 1. In the tight-binding
approximation, the Hamiltonian is described as \cite{r21aa}

\begin{eqnarray}
H &=& \sum_{n}[\epsilon_{n}c^{\dag}_{n}
c_{n}-t_{n}(c^{\dag}_{n} c_{n+1}+H.c.)]\nonumber \\
&&+\sum_{n=1}^{N}[(\epsilon_{bn}+V_{g})b^{\dag}_{n}b_{n}-t_{bn}(c^{\dag}_{n}
b_{n}+H.c.)],
\end{eqnarray}
where each lattice site represents a poly(dA)-poly(dT) base or
poly(dG)-poly(dC) base for $n \in [1,N]$. The operator
$c^{\dag}_{n}(c_{n})$ creates(annihilates) a charge at the $n$th
site of the base pair in the DNA chain ($1 \leq n \leq N$), of the
left electrode ($n \leq 0$),  of the right electrode ($n \geq
N+1$). The operator $b^{\dag}_{n}(b_{n})$ creates(annihilates) a
charge at the $n$th site of the substrate ($1 \leq n \leq N$).
$\epsilon_{n}$ and $t_{n}$, respectively, denote the site energy
and hopping integral in DNA chain, $\epsilon_{bn}$ is the site
energy in the substrate, $t_{bn}$ stands for coupling between
chain and substrate, and $V_{g}$ is the gate-voltage.

On each substrate site there is a highly localized orbit formed
via the interaction of the base pair, the sugar-phosphate
structure and the substrate surface. So, it is reasonable to
consider that different base pairs corresponds to different values
of coupling strength $t_{bn}$ and side site energy
$\epsilon_{bn}$, that is, for base pair $GC$, we have
$\epsilon_{bn}=\epsilon_{bGC}$, $t_{bn}=t_{bGC}$; for base pair
$AT$, we have $\epsilon_{bn}=\epsilon_{bAT}$, $t_{bn}=t_{bAT}$.
Obviously, this is a kind of longitudinal short-range correlation
analogous to a layer structure that may produce a band of extended
states \cite{r23}.

If the base pair GC (or AT) is distributed randomly and
independently with a probability $p$ (or $1-p$), the disorder
strength W will depend on $p$ and $|\epsilon_{AT}-\epsilon_{GC}|$.
For $p=1$ or $0$ cases, we have a period chain with two energy
bands, the highest occupied molecular orbital (HOMO) and the
lowest unoccupied molecular orbital (LUMO). For a fixed
$|\epsilon_{AT}-\epsilon_{GC}|\neq 0$ case, the most disordered
state is given by $p=0.5$ \cite{r23} and in this case the
transmittivity of the LUMO is almost zero. So the following text
only focuses on the HOMO with $p=0.5$.

%%%%%%%%%%%%%  figure 1

\begin{figure*}[htb]
  \centering
  \includegraphics[width=17cm]{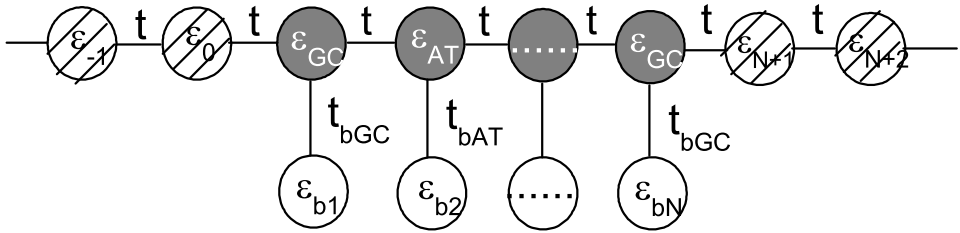}
  \caption{Schematic illustration of extending a DNA chain with
substrate to a very long chain. The black circles represent the
DNA chain, white circles denote substrate, and shaded circles
stand for the attached leads. Electronic pathways are shown as
lines.}
\end{figure*}

%%%%%%%%%%%%%  figure 1

In the site representation, the Schr\"{o}dinger equation $H|\psi
\rangle = E |\psi \rangle$ corresponding to Eq. (1) becomes

\begin{eqnarray}
(E-\epsilon_{n})\psi_{D,n} = -t_{n-1}\psi_{D,n-1}-t_{n}\psi_{D,n+1} -t_{bn} \psi_{S,n},\nonumber\\
(E-\epsilon_{bn}-V_{g})\psi_{S,n} = -t_{bn}\psi_{D,n}.
\end{eqnarray}
Here, $\psi_{D,n}$ is the amplitude of the wave function at the
$n$th site of the chain(the $n$th base pair site) and $\psi_{S,n}$
is the amplitude of the wave function at the $n$th site of the
substrate. After the decimation of $\psi_{S,n}$\cite{r21,r22}, the
corresponding Schr\"{o}dinger equation of the Hamiltonian (1) is

\begin{eqnarray}
 \label{seq1}
(\epsilon^{\prime}_{n}-E)\psi_{D,n} =
t_{n-1}\psi_{D,n-1}+t_{n}\psi_{D,n+1}.
\end{eqnarray}
Here, $\epsilon^{\prime}_{n}$ is the renormalized on site energy,
which is defined as
\begin{eqnarray}
 \label{onsite1}
\epsilon^{\prime}_{n} =
\epsilon_{n}+\frac{t^{2}_{bn}}{E-(\epsilon_{bn}+V_{g})}.
\end{eqnarray}
The renormalized site energy $\epsilon^{\prime}$ includes the
influence of substrate on conduction state in DNA chain, and it is
a function of the gate voltage $V_{g}$ and the incident electron
energy $E$ \cite{r15}.

The Eq. (\ref{seq1}) can be rewritten as relations between
coefficient of adjacent pints in the form of the transfer matrix

\begin{equation}
 {\psi_{D,n+1} \choose \psi_{D,n}}=T_{n}{\psi_{D,n} \choose
\psi_{D,n-1}},
\end{equation}
with
\[T_{n}=\left(\begin{array}{r@{\quad \quad}l}\frac{(\epsilon^{\prime}_{n}-E)}{t_{n}}
   & -\frac{t_{n-1}}{t_{n}}
\\1&  0\end{array}\right).\]
For a 1-D chain with length $L$, the coefficients at one end are
related to the coefficients at the other end with the transfer
matrices
\begin{equation}
 \label{transfer1}
 \left( \begin{array}{c} \psi_{L} \\ \psi_{L-1} \end{array} \right) = \left( \prod_{j=1}^{L-1}
 T_{L-j} \right)
 \left( \begin{array}{c} \psi_{1}
 \\ \psi_{0} \end{array} \right).
 \end{equation}
The Lyapunov exponents(LE) are the logarithms of eigenvalues of
the transfer matrix, and can be calculated by using the
transfer-matrix method, in which the orthonormalization procedure
is adopted \cite{sec22}.

In this paper, the transmittivity \cite{r29} and the localization
length (LL) \cite{r29,r28} which is the inverse of the LE are
calculated with the transfer matrix (TM). The current is computed
with the Landauer-B\"{u}ttiker formula

\begin{equation}
 \label{current1}
 I = \frac{2e}{h}\int^{+\infty}_{-\infty}T(E)(f_{L}-f_{R})dE ,
\end{equation}
here $f_{L/R}(E)=\{1+exp[(E\pm eV_{d}/2-E_{F})/k_{B}T]\}^{-1}$ is
the Fermi distribution on the left/right lead. $E_{F}$ and $V_{d}$
are the equilibrium Fermi energy and applied voltage,
respectively. $T(E)$ is the transmission through N base pairs.

\section{NUMERICAL RESULTS}

In the numerical calculations we adopt a DNA chain with different
length $N$. The lattice structure is illustrated in Fig. 1. The
values of the parameters are chosen as the following \cite{r9}:
$\epsilon_{n}$ is the average of the complementary bases energies,
$\epsilon_{GC}=(\epsilon_{G}+\epsilon_{C})/2$,
$\epsilon_{AT}=(\epsilon_{A}+\epsilon_{T})/2$\cite{r24}. Here,
$\epsilon_{G}=7.75eV$, $\epsilon_{C}=8.87eV$,
$\epsilon_{A}=8.24eV$, $\epsilon_{T}=9.14eV$. The site energy on
the lead $\epsilon_{m}=5.36$ is the work function of platinum. The
site energy $\epsilon_{bGC}=10.1eV$ and $\epsilon_{bAT}=11eV$ are
slightly higher than the base pair energy $\epsilon_{n}$. For
simplicity, both the hopping integral on the DNA and at the
contacts are fixed at $0.4eV$, to minimize the contact scattering
\cite{r25}. The hopping integral on the leads is $12eV$. The
surface coupling parameters $t_{bGC}=4.4eV$ and $t_{bAT}=5.0eV$
which are stronger than the usual coupling between base pair and
sugar-phosphate structure \cite{r16,r18}. The latter is about
$1.0eV$\cite{r9,r26} and is neglected here. Fermi energy is set to
be $5.36 eV$ through out the paper.

%%%%%%%%%%%%%%%%% figure 2

\begin{figure}[htb]
  \centering
  \includegraphics[width=8.5cm]{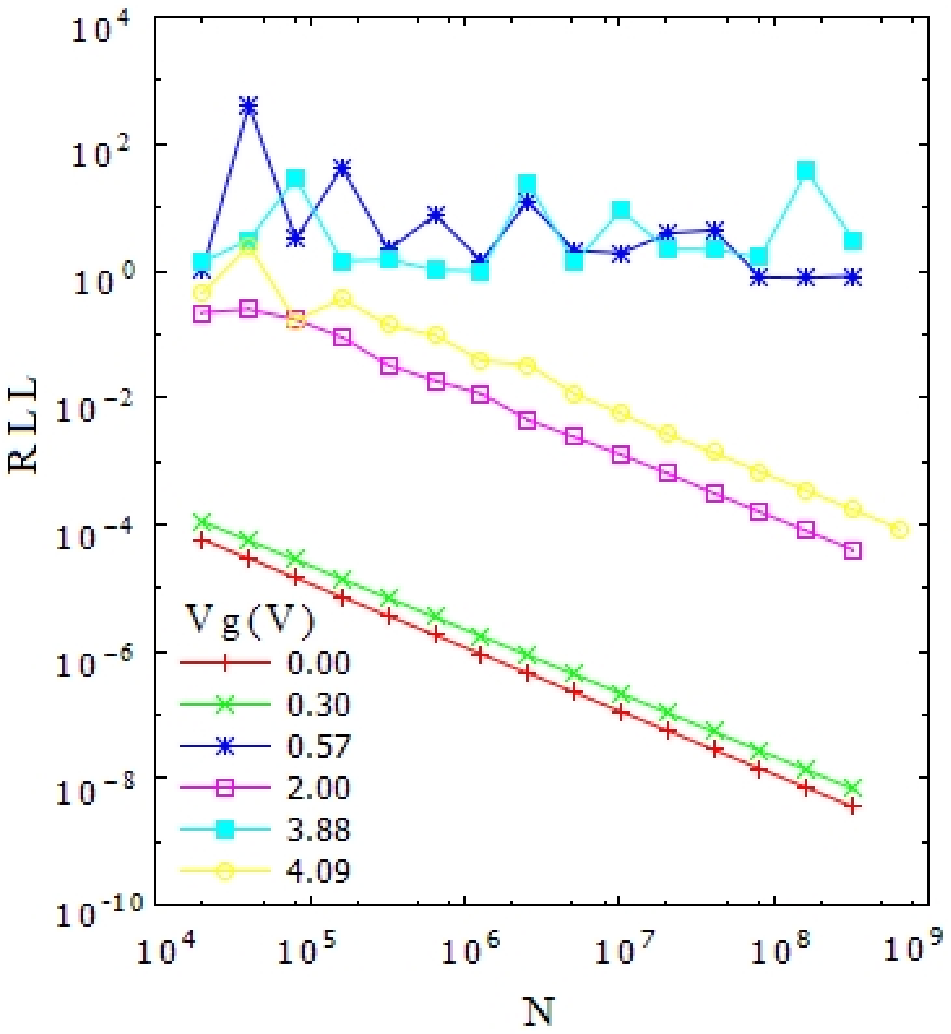}
  \caption{(Color online.) The rescaled localization length (RLL)  as a function of
chain length $N$ at $E=E_{F}$; red line for $V_{g}=0.00V $, green
line for $V_{g}=0.30V $, blue line for $V_{g}=V_{g1}\approx
0.57V$; pink line for $V_{g}=2.00V $, light blue line for
$V_{g}=V_{g2}\approx 3.88V$; yellow line for $V_{g}= 4.09V$.}
\end{figure}

%%%%%%%%%%%%%%%%%%  figure 2

%%%%%%%%%%%%%%%%%%  figure 3

\begin{figure}[htb]
  \centering
  \includegraphics[width=8.5cm]{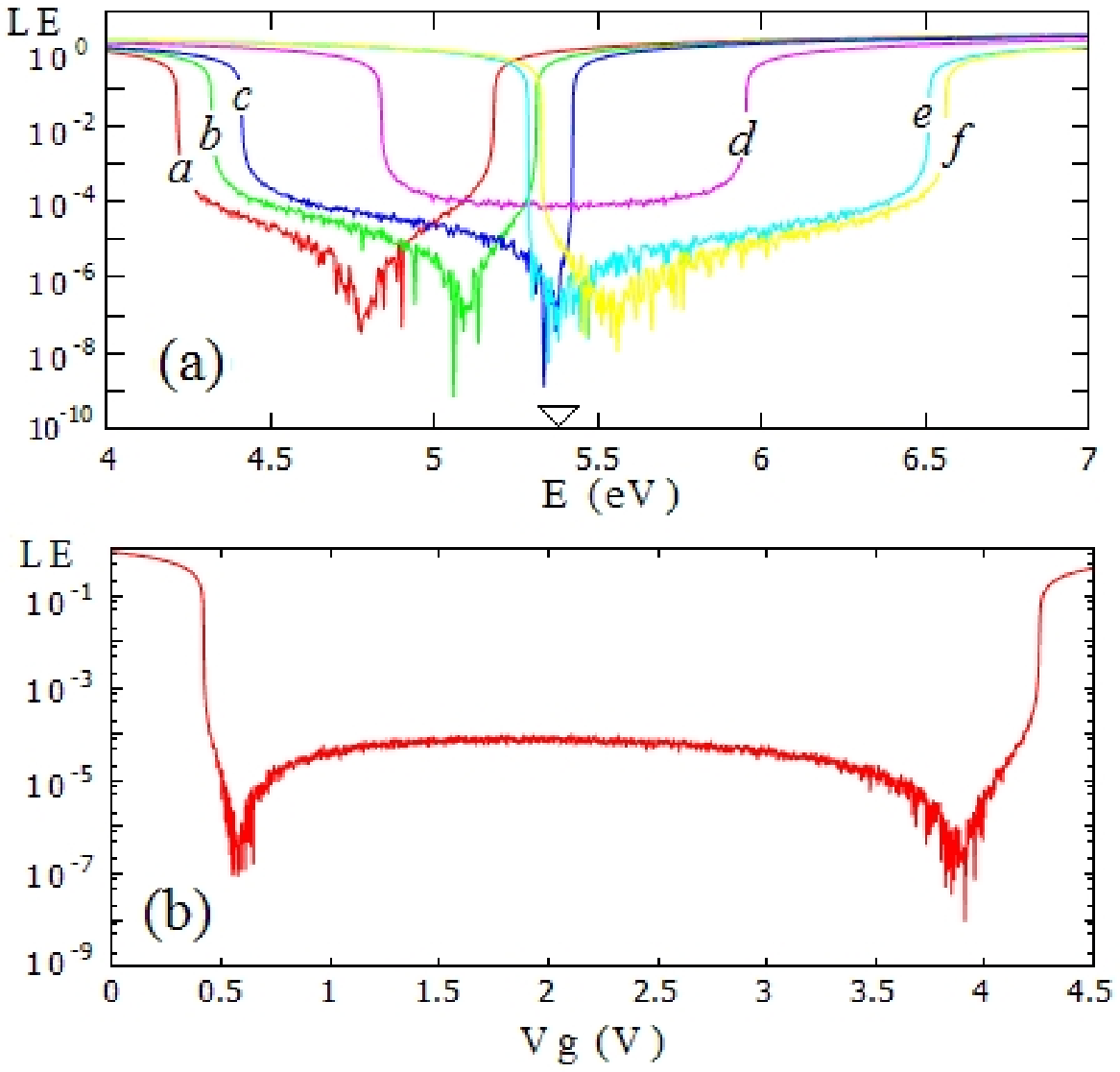}
  \caption{(Color online.) (a) Lyapunov Exponent(LE) as a function of energy E with
different $V_{g}$; curve a, $V_{g}=0 V$; b, $V_{g}=0.30V $; c,
$V_{g}= V_{g1}\approx 0.57V $; d, $V_{g}=2.00V $; e, $V_{g}=
V_{g2}\approx 3.88V $; f, $V_{g}=4.09V$. The open triangle marks
$E_{F}=5.36eV$. (b) LE versus $V_{g}$ at $E_{F}$. The DNA length N
is $10^{6}$.}
\end{figure}

%%%%%%%%%%%%%%%%%%  figure 3

First, we present the results for a DNA chain with different
length $N$. Figure 2 shows the rescaled localization length(RLL)
$\Lambda/N$ as a function of chain length $N$ with different gate
voltage $V_{g}$ at Fermi energy. Here, $\Lambda$ represents the
localization length. It can be seen, for most values of $V_{g}$,
that the $\Lambda/N$ decreases with increasing $N$, while, for
some given gate voltages $V_{g1} \sim 0.57 V$ and $V_{g2} \sim
3.88 V$, $\Lambda/N$ is slightly increased with $N$. The RLL
$\Lambda/N$ decays with power rate as increasing N for localized
state, while it fluctuates throughout for extended state
\cite{r32,r33}. This difference in scaling behavior indicates that
there may exist delocalized states for given gate-voltages in the
thermodynamic limit.

In Fig. 3(a), we plot the Lyapunov exponent as a function of the
energy $E$ with different gate voltage $V_{g}$ for $N=10^{6}$. It
can be seen, for all the investigated values of $V_{g}$, that the
smallest Lyapunov exponent always lies in a narrow region of
energy and appears at different energy $E$. Since the transport
properties of the system mainly depends on the quasi-particle's
states near the Fermi levels, Fig. 3(b) shows Lyapunov exponent as
a function of the gate voltage $V_{g}$ at Fermi energy. It can be
seen that there exist two valley-like structures at
$V_{g}=V_{g1}\sim 0.57$ and $V_{g}=V_{g2}\sim 3.88$, which are
exactly the same gate voltages that corresponds to the
delocalization behaviors in Fig. 2. This valley-like structure
indicates that there may exist a delocalization-localization
translation around the valleys. Note that only the points at the
bottom of the valleys correspond to the smallest Lyapunov
exponent. Combined the results of Fig. 3(a) and 3(b), we could
see, at the bottom of the valleys, i.e. $V_{g1}$ and $V_{g2}$,
that the narrow energy region which corresponds to the smallest
Lyapunov exponent appears around Fermi levels. Considering the
scaling behavior of the RLL in Fig. 2, we could conclude that
these states, which correspond to the smallest Lyapunov exponent
at $V_{g}=V_{g1}$ and $V_{g}=V_{g2}$ in Fig. 3(b), are isolated
extended states at Fermi energy.

In order to understand the origin of these isolated extended
states, Eq. (\ref{seq1}) should be discussed further. Since only
two kinds of base pairs G-C and A-T are considered in present
system, there are two kinds of $\epsilon^{\prime}_{n}$, i.e.
$\epsilon^{\prime}_{AT}$ and $\epsilon^{\prime}_{GC}$. If
parameters satisfy the following conditions,

\begin{eqnarray}
\label{condition1} \epsilon^{\prime}_{AT} = \epsilon^{\prime}_{GC}
= \epsilon^{\prime},
\end{eqnarray}
\begin{eqnarray}
\label{condition2} E \in [\epsilon^{\prime}-2t,
\epsilon^{\prime}+2t],
\end{eqnarray}
then eq. (\ref{seq1}) becomes
$$(\epsilon^{\prime}-E)\psi_{D,n} = t\psi_{D,n-1}+t\psi_{D,n+1}$$
which corresponds to an Schr\"{o}dinger equation for a periodic
chain. In other words, the solution of the above conditions
corresponds an extended state with energy $E$. So Eq.
(\ref{condition1}) and (\ref{condition2}) are a criteria for
finding an isolated extended states within our model. The origin
of this extended state comes from the constructive interference
between DNA chain sites and substrate sites \cite{r30}. Condition
for Eq. (\ref{condition2}) ensures E in the propagation band of
the renormalized period chain with onsite energy
$\epsilon^{\prime}$. Here, $\epsilon^{\prime}$ varies with E
according to Eq. (\ref{condition1}). Both Eq. (\ref{condition1})
and (\ref{condition2}) are independent of GC concentration $p$ and
base pair arrangements, which indicates that the existence of the
extended state will not be impacted by specific base pair
sequences. This mechanism is similar to the cases in refs.
\cite{r23,r31}.

Substitute the renormalized on site energy $\epsilon^{\prime}$ in
eq. (\ref{onsite1}) into the above criteria for isolated extended
states, Eq. (\ref{condition1}) and (\ref{condition2}) turn into
the following forms:
\begin{equation}
\label{condition11}
 E -eV_{g} = -\frac{\alpha}{2}\pm
\sqrt{\gamma}, (\gamma \geq 0).
\end{equation}
\begin{equation}
E \in [\epsilon^{\prime}_{+}-2t, \epsilon^{\prime}_{+}+2t]\cup
[\epsilon^{\prime}_{-}-2t, \epsilon^{\prime}_{-}+2t].
\end{equation}
Here, $\triangle \epsilon = \epsilon_{AT}-\epsilon_{GC}$, $\alpha
= \triangle
\epsilon^{-1}(t^{2}_{bAT}-t^{2}_{bGC})-(\epsilon_{bAT}+\epsilon_{bGC})$,
$\beta = \epsilon_{bAT}\epsilon_{bGC}[\triangle
\epsilon^{-1}(t^{2}_{bAT}\epsilon^{-1}_{bAT}-t^{2}_{bGC}\epsilon^{-1}_{bGC})-1]$,
$\gamma = \beta+\alpha^{2}/4$, and
$\epsilon^{\prime}_{\pm}=\epsilon_{AT}+t^{2}_{bAT}/(-\alpha/2\pm\sqrt{\gamma}-\epsilon_{bAT})$.
Therefore, the relation of the isolated extended state energy E
and the gate voltage $V_{g}$ is two pieces of paralleled line
segments with slope $e$ and with the terminal vertexes depending
on site energies and hopping integrals. When $E=E_{F}=5.36eV$, Eq.
(\ref{condition11}) gives two solutions for $V_{g}$: $V_{g}\approx
0.57 (=V_{g1})$ and $V_{g}\approx 3.88  (=V_{g2})$, which means,
for these two cases, that there is an isolated extended state at
Fermi level $E_{F}$. These results are consistent with the
previous results in Fig. (2) and Fig. (3).

Figure 4 shows the transmission coefficients as a function of
energy for different $V_{g}$ and $N$. It can be seen that there is
a transmission band for the chain with different gate voltages
when the chain length is short. With the increase of chain length
$N$, some transmission band would disappear, i.e. for $V_{g}=2$
case, while others may leave a sharp peak around the value of
energy of the isolated extended state, i.e. for $V_{g}=0$,
$V_{g1}$, and $V_{g2}$ cases. This implies that there are two
kinds of mechanisms for transmission. The former depends on the
effectively delocalized states\cite{r34}, which forms in a system
shorter than its localization length, while the latter depends on
an isolated extended state. With the increase of system length,
the effectively delocalized states disappear, however, the
isolated extended state is still there. In other words, with the
increase of N, the transmission band decays but leaves a sharp
peak around the energy of an isolated extended state for
$V_{g}=0$, $V_{g1}$, and $V_{g2}$ cases, or vanishes for $V_{g}=2$
case.

In Fig. 5, we plot the conductance of chain with different gate
voltage as a function of chain length $N$. It can be seen that the
conductance $G$ decreases rapidly with the increase of the length
$N$ for cases with $V_{g} \neq V_{g1} or V_{g2}$. This implies
that the number of the effectively delocalized states around Fermi
energy for such cases decreases rapidly with the increase of
length of the system. It can also be seen that, for $V_{g} =
V_{g1}$ and $V_{g} = V_{g2}$ cases, the conductance shows a strong
oscillation. The oscillation might be caused by the coherent
scattering of Bloch's wave at the two contacts of the leads
\cite{r25}. For $V_{g}=4.085$ case, $G_{0}$ decays exponentially
with some oscillation because the Fermi level is close to the
isolated extended state. This shows that conductance mainly
depends on the chain length $N$ and the energy position of the
isolate extent states.

%%%%%%%%%%%%%%%%% figure 4

\begin{figure}[htb]
\centering
\includegraphics[width=8.5cm]{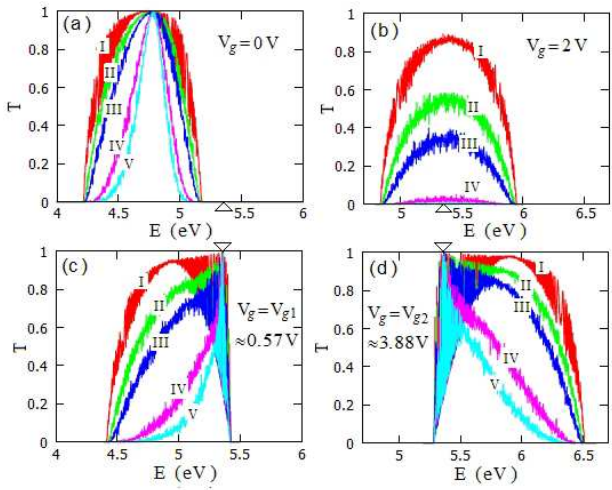}
\caption{(Color online.) Transmission coefficient $T(E)$ of a
chain with different $V_{g}$ as a function of energy. The curve I,
II, III, IV, V represent for the cases with $N=1000$, $N=5000$,
$N=10000$, $N=50000$, and $N=100000$, respectively. The
probability $p= 0.5$. The results are averaged over 100 samples.
The Fermi level $E_{F}=5.36eV$ is labelled by the open triangles.}
\end{figure}

%%%%%%%%%%%%%%%%%%  figure 4

%%%%%%%%%%%%%%%%% figure 5

\begin{figure}[htb]
\centering
\includegraphics[width=8.5cm]{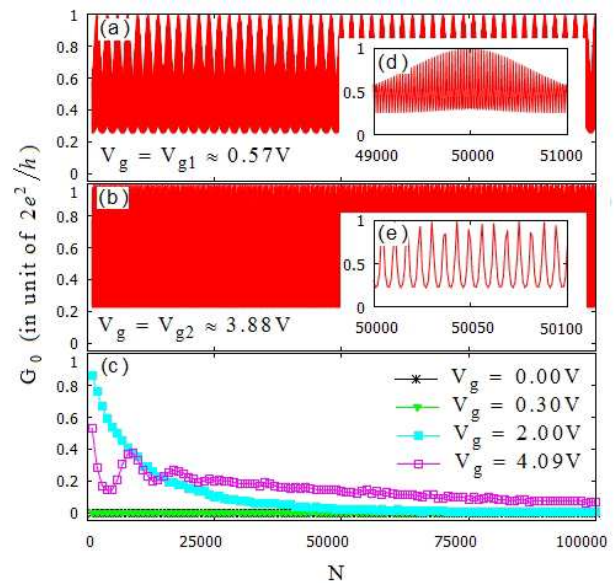}
\caption{(Color online.) Conductance  $G$ as a function of DNA
length N with (a) $V_{g}=V_{g1}$, (b) $V_{g}=V_{g2}$, (c)
$V_{g}=0eV$, $0.30eV$, $2.00eV$, and $4.09eV$. The inset Fig. (d)
and (e) show the details of the oscillation in figure (a) and (b),
respectively. The oscillation are very dense in figure (a) and
(b). The results in figure (c) are averaged over 1000 samples.}
\end{figure}

%%%%%%%%%%%%%%%%%%  figure 5

%%%%%%%%%%%%%%%%% figure 6

\begin{figure}[htb]
\centering
\includegraphics[width=8.5cm]{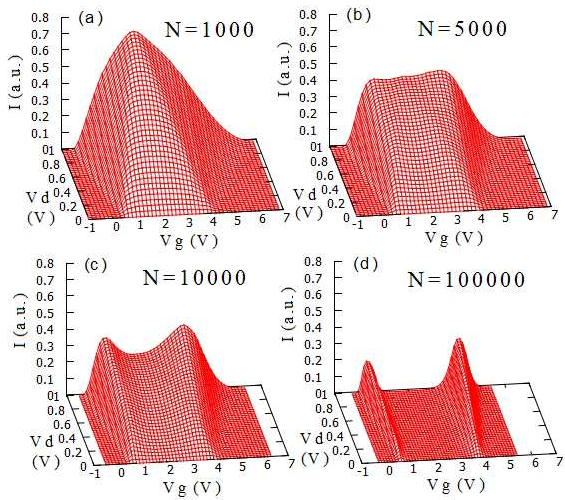}
\caption{(Color online.) Total current$I$ of a chain with
different length N as a function of the gate voltage $V_{g}$ and
applied voltage $V_{d}$ on the two leads with (a) $N=1000$, (b)
$N=5000$, and (c) $N=10000$; (d) $N=100000$. The temperature is
0.1 K. The results are averaged over 100 samples. Other parameters
are the same as those in Fig. 2.}
\end{figure}

%%%%%%%%%%%%%%%%%%  figure 6

In order to consider the combined effect of the gate voltage and
the chain length, in Fig. 6, we plot $I-V_{d}$ characters of DNA
with different length $N$ for different values of gate voltage
$V_{g}$. The current $I$ is averaged with 100 realizations. It can
be seen, with the increase of the length $N$, that, for some kind
of $V_{g}$, the current $I$ decreases and even becomes zero for
some certain values of gate voltage, while, for the other kind of
$V_{g}$ around $V_{g1}$ and $V_{g2}$, the current still exists for
the regions of $V_{g}$ around $V_{g1}$ and $V_{g2}$ when $N$ is up
to $10^{5}$. This shows that transport properties of DNA are
suppressed by the length.

The difference between these behaviors of currents lies in
different conduction mechanisms and can be understood by the
concept of the effective delocalized state\cite{r34}. For the
short chain with $N \leq \Lambda$ cases, the quasiparticles are in
the effectively delocalized states with high transmittivity spread
all over the band and conceal the impact of the isolated extended
state. This situation corresponds to the current peak for the case
of $V_{g}=2$ in Fig. 6. With further increases of N, the
effectively delocalized states eventually vanish, while the impact
of the isolated extended state emerges. Therefore, two current
peak bands form respectively around $V_{g1}$ and $V_{g2}$ where
the isolated extended state lies around the Fermi level. However,
the two peaks are not exactly at $V_{g1}$ and $V_{g2}$, because
the Landauer-B\"{u}ttiker formula is asymptotically equal to $I =
2eh^{-1}\int^{E_{F}+eV_{d}/2}_{E_{F}-eV_{d}/2}T(E)dE$ at low
temperature, and the maximum of current does not require the
transmission peak exactly at $E_{F}$, but the maximum area under
the curve of $T(E)$ in the conducting interval
$[E_{F}-eV_{d}/2,E_{F}+eV_{d}/2]$. However, the larger the length
N is, the more proximity to $V_{g1}$ and $V_{g2}$ the current
peaks get. In other words, the effective delocalized states will
vanish with the increase of chain length, while the isolated
extended states may still exist.

\section{CONCLUSIONS AND DISCUSSION}

We have investigated the coherent transport properties of a DNA
chain with the interaction between DNA and substrate. Using the
tight-binding model and transfer matrix technique, it is found
that there is an isolated extended state which is robust against
the DNA sequence rearrangements. Physically, the origin of such an
isolated extended state is the interference between DNA chain
sites and substrate sites. Based on this mechanism, the
controllable long distance coherent transport in mesoscale can be
achieved by fabricating single molecular FET with long disordered
DNA chains (e.g., $\lambda-DNA$) adhering to substrate surface.
This mechanism is also valid for the general nanowire of binary
alloy or conductive polymer interacting with substrate surface.
Furthermore, since the principle of complementary nucleobases is a
natural transverse correlation of short range, this mechanism may
be generalized to the ladder model of DNA, and may therefore
impact the spin selective transport through double strand DNA as
well\cite{r35}, which needs further study of both theory and
experiment. The conductance and current-bias characteristics shows
that Metal-insulator transition (MIT) could emerge by varying the
gate voltage or the chain length. Two different conduction
mechanisms attributing to effectively delocalized states and
isolated extended states, respectively, are proposed. These
results may provide perspectives for experimental work aimed at
controlling charge transport through DNA-based nanodevices.

\section{Acknowledgements}

We would like to thank Professor S. J. Xiong for many helpful
discussions. This work was supported by the National Fund of
Natural Sciences of China (Grant No. 11204372) and a grant for
STRP from the HNED of China (Grant No. 14B140013).

\end{document}